\documentclass[aps,twocolumn,superscriptaddress]{revtex4}

\usepackage{amsmath,amssymb,graphicx}
\usepackage{algorithmic}
\usepackage{enumerate}
\usepackage{color}

\graphicspath{{./fig-jpg/}{./fig-ps/}}

\begin{document}

\title{Adiabatic invariant analysis of dark and dark-bright soliton stripes 
in two-dimensional Bose-Einstein condensates}

\author{P. G. Kevrekidis}
\affiliation{Department of Mathematics and Statistics, University of Massachusetts,
Amherst, Massachusetts 01003-4515 USA}

\author{Wenlong Wang}
\email{wenlongcmp@gmail.com}
\affiliation{Department of Theoretical Physics, Royal Institute of Technology, Stockholm, SE-106 91, Sweden}

\author{R. Carretero-Gonz{\'a}lez}
\affiliation{Nonlinear Dynamical Systems
Group,\footnote{\texttt{URL}: http://nlds.sdsu.edu}
Computational Sciences Research Center, and
Department of Mathematics and Statistics,
San Diego State University, San Diego, California 92182-7720, USA}

\author{D. J. Frantzeskakis}
\affiliation{Department of Physics, National and Kapodistrian University of Athens,
Panepistimiopolis, Zografos, 15784 Athens, Greece}

\begin{abstract}
  In the present work, we develop an adiabatic invariant approach for
  the evolution of quasi-one-dimensional (stripe) solitons 
  embedded in a two-dimensional Bose-Einstein condensate.
  The results of the theory are obtained both for the 
  one-component case of dark soliton stripes,
  as well as for the considerably more involved case of
  the two-component dark-bright (alias ``filled dark'') soliton stripes. 
  In both cases, analytical predictions regarding the stability and 
  dynamics of these structures are obtained. 
  One of our main findings is the determination of the instability
  modes of the waves as a function of the parameters of the system
  (such as the trap strength and the chemical potential). Our analytical 
  predictions are favorably compared with results of direct numerical simulations. 
\end{abstract}

\maketitle

\section{Introduction}

A theme of wide interest over the last two decades is the study
of dark solitons; relevant explorations were physically motivated
in nonlinear optics~\cite{Kivshar-LutherDavies} and more recently
have been broadly extended to atomic Bose-Einstein condensates (BECs)~\cite{djf}.
One of their two-dimensional (2D) generalizations, i.e., 
vortices ---which play 
a prominent role in  
nonlinear field theory~\cite{Pismen1999}--- 
have also attracted attention in nonlinear optics~\cite{YSKPiO,desa}
and atomic BECs~\cite{fetter1,fetter2}.
These two structures are intimately connected 
through their topological nature: vortices can be thought of
as a 2D ``incarnation'' of a dark soliton ---possessing
a $2\pi$ phase winding. However, there is also an important link
from the point of view of stability analysis, namely dark
solitons become unstable in higher dimensions~\cite{kuzne,kidep},
giving indeed rise to vortices. The relevant dynamics is characterized 
by the manifestation of the so-called transverse (or ``snaking'') instability, 
which leads to the undulation and the eventual breakup of dark solitons 
into multi-vortex patterns.
This feature has been used experimentally since early on as a means
of producing vortices, both in optics~\cite{tikh} and in BECs~\cite{watching}, 
and has been a subject of continuing theoretical interest~\cite{smirnov,hoefer,ai}.
Mechanisms on how to avoid the instability have also been explored (see, e.g., Ref.~\cite{us}).

In the recent work of Ref.~\cite{ai}, we developed an approach to tackle
transverse instabilities, with a special emphasis on the case examples 
of ring dark solitons (studied in optics~\cite{kivyang,rings,rings1} 
and BECs~\cite{rings2,rings3,rings4})
and spherical shell solitons (also of wide interest in
the same areas~\cite{kivyang,carr,wenlong,hau}). The technique
was based on a generalization of the adiabatic invariant
(or so-called ``Landau dynamics'') approach. This was
a technique earlier utilized for dark solitons in 
one-dimensional (1D) settings~\cite{konotop_pit,konotop_pit2} and for 
ring dark solitons in quasi-1D ones~\cite{korneev}.

Our scope in the present work is to extend the relevant considerations 
to the case of the dark soliton stripe for the one-component 
case, as well as the dark-bright (alias ``filled dark'') soliton stripe in the case
of two-component systems of the nonlinear Schr{\"o}dinger (NLS) type.
We develop, in both cases, the adiabatic invariant theory ---extending it 
in this way to the multi-component, multi-dimensional case--- and derive 
the equations of motion of these ``solitonic filaments'', in the 
presence of curvature, as well as in that of the external potential 
relevant to BECs. Subsequently, 
from these 1D partial differential equations (PDEs) characterizing
the $x$-position of the filament as a function of $(y,t)$, assuming that 
the filament extends along the $y$-direction,
we infer the equilibrium states, i.e., the homogeneous equilibria
corresponding to straight filaments. We linearize around these equilibria
to identify their modes of potential instability and their
corresponding wavenumbers as a function of parameters, such as
the chemical potential of the system. Finally, we test
all of the above existence, stability, and dynamical predictions
against numerical simulations, finding good agreement with the
corresponding PDE results (both analytical ones ---e.g., for the
linearization--- and lower-dimensional, effective dynamical ones).

Our presentation is structured as follows. First, we give a summary
of our analytical results both for the single- and for the two-component
case. Then, we proceed to test the conclusions of our analysis against
the stability analysis and dynamics of the original, full 2D, PDE. 
Finally, we summarize our findings and present a number of
possibilities for future work.

\section{Mathematical Formulation and Analytical Results}

\subsection{One-Component Case}

Our starting point is the dimensionless 1D NLS equation
---also referred to as the Gross-Pitaevskii (GP) equation--- 
which includes the external trapping potential $V(x)$, appearing 
generically in the BEC context; the equation is 
of the form:
\begin{eqnarray}
  i u_t =-\frac{1}{2} u_{xx} + |u|^2 u + V(x) u.
  \label{dark1}
\end{eqnarray}
For the derivation of both the 1D and 2D models used herein in
their dimensionless form (from their dimensional variants),
the reader can consult, e.g., Ref.~\cite{rcg:BEC_book2}.
In the absence of external potential, $V(x)=0$, and for
a background density (equal to the chemical potential) $\mu$, 
the conserved energy assumes the form: 
\begin{eqnarray}
  H_{\rm 1D}=\frac{1}{2}  \int_{-\infty}^{\infty} |u_x|^2 +
  \left(|u|^2-\mu \right)^2 dx.
\notag
\end{eqnarray}
In the same case (where the potential is absent), Eq.~(\ref{dark1}) 
possesses a dark soliton solution, of 
position $\xi$ and velocity $v=d\xi/dt \equiv \dot{\xi}$, given by: 
\begin{eqnarray}
u(x,t)=e^{-i \mu t} \left[
  \beta \tanh \left(\beta
(x-{\xi}) \right) + i v \right],
\label{dark2a}
\end{eqnarray}
with $\beta=\sqrt{\mu - v^2}$. 
For this solution, the energy yields: $H_{\rm 1D}=(4/3) (\mu-\dot{\xi}^2)^{3/2}$.
We then follow Refs.~\cite{konotop_pit,konotop_pit2}
and use this energy as an 
{\em adiabatic invariant} (AI) ---i.e., an invariant under slow variations---
in the presence of a slowly-varying potential $V(x)$.
This is justified by the consideration that, in this case, 
the background density 
$\mu$ will be {\em slowly-varying} according to the transformation 
$\mu \rightarrow \mu - V(x)$. 
Therefore, assuming the AI of this quantity, we obtain
\begin{eqnarray}
  H_{\rm 1D} = \frac{4}{3} \left(\mu - V({\xi}) -\dot{\xi}^2 \right)^{3/2},
  \label{dark3H}
\end{eqnarray}
which gives, after taking a time derivative, the following
equation of motion for the dark soliton position:
\begin{eqnarray}
  \ddot{\xi}=- \frac{1}{2} V'({\xi}).
  \label{dark3}
\end{eqnarray}
This result, obtained originally
in Ref.~\cite{busch} and retrieved in Ref.~\cite{konotop_pit}, 
is well-known to be in very good agreement with numerical results
for large $\mu$~\cite{konotop_pit,konotop_pit2,rcg:BEC_book2}.
In this limit, the dark solitons 
can be thought of as particles
bearing no internal structure, enabling the application of
this effective particle theory.

Our considerations are geared towards generalizing the above
ideas to 2D. Let us then consider the 2D NLS equation:
  \begin{eqnarray}
    i u_t =-\frac{1}{2} \left(u_{xx}+u_{yy}\right) + |u|^2 u + V(x) u,
    \label{extr}
  \end{eqnarray}
where, importantly, we consider the case $V=V(x)$ corresponding to
only trapping along the (longitudinal) $x$-direction.
This 2D NLS 
conserves the 
Hamiltonian:
\begin{eqnarray}
  H_{\rm 2D}=\frac{1}{2}  \iint_{-\infty}^{\infty} \left[ |u_x|^2 + |u_y|^2 +
  \left(|u|^2-\mu \right)^2 \right] dx\, dy.
\notag
\end{eqnarray}

Now, assuming an ansatz of the form (\ref{dark2a}) with
the center position $\xi$ not solely a function of $t$, but also
a function of the transverse variable $y$, i.e., $\xi={\xi}(y,t)$,
we are able to describe solutions of the form of 
a dark soliton stripe,
or {\em soliton filament}, that runs along the $y$-direction.
Evaluating the 2D Hamiltonian for this dark soliton stripe yields
an ``effective energy'' (an AI again) of the form:
\begin{eqnarray}
  E= \frac{4}{3} \int_{-\infty}^{\infty} \left(1 + \frac{1}{2}{{\xi}}_y^2 \right)
  \left(\mu - V({\xi}) -{{\xi}}_t^2 \right)^{3/2}  dy.
  \label{dark5}
\end{eqnarray}
Here, the transverse energy contribution (corresponding to the $|u_y|^2$ term) has
been accounted for through the term proportional to ${{\xi}}_y^2$.
One can try to obtain various pieces of quantitative information based
on this ``effective Hamiltonian'' describing the {\em transverse
motion of the soliton filament}. 
Similarly to the 1D case, we take 
$dE/dt=0$ and integrating 
by parts along the $y$-direction (and considering 
localization of the
solution along the $y$-direction, so that partial derivatives with respect to $y$ at
$y=\pm\infty$ are zero), we obtain the following effective PDE for the dark soliton 
filament's dynamical evolution:
\begin{eqnarray}
  {{\xi}}_{tt} B + \frac{1}{3} {{\xi}}_{yy} A = {{\xi}}_y\, {{\xi}}_t\, {{\xi}}_{yt}
  -\frac{1}{2} V'({\xi}) \left( B - {{\xi}}_y^2 \right),
\label{1D_PDE_DS}
\end{eqnarray}
where $A=\mu - V({\xi}) - {{\xi}}_t^2$ and $B= 1 + \frac{1}{2} {{{\xi}}_y^2}$.
One can then make the following relevant observations regarding this novel
emerging PDE model:

\begin{itemize}
\item[(i)]
For weak undulations, and in the absence of the potential, the dynamics
is described by
\begin{eqnarray}
{{\xi}}_{tt}  + \frac{1}{3}\mu\, {{\xi}}_{yy} =0,
\notag
\end{eqnarray}
yielding the proper linear growth rate of the transverse
instability~\cite{kuzne}. Note that such an instability for 
dark solitons is {\em only} present in the elliptic 
dispersion case 
[dispersion term equals to $\frac{1}{2}(u_{xx}+u_{yy})$, as in the case under 
consideration]. 

\item[(ii)]
Assuming that ${\xi}={\xi}(t)$ is only a function of time yields
\begin{eqnarray}
{{\xi}}_{tt} = -\frac{1}{2} V'({\xi}),
\notag
\end{eqnarray}
i.e., Eq.~(\ref{dark3}) is recovered. 

\item[(iii)]
It is possible to obtain existence and stability information for the
dark soliton stripe. A particularly interesting example, even at the linear setting,
concerns the case with the ---generic for BECs--- 1D parabolic trap  
 $V(x)=\frac{1}{2} \Omega^2 x^2$.
This case concerns a 1D dark-soliton
embedded in a longitudinal
trap, while the transverse 
direction remains untrapped. Naturally, $\xi(y,t)=0$ is the stationary state, 
corresponding to a dark soliton stripe 
located at the potential minimum. Applying the normal mode ansatz
${\xi}(y,t)=X_0 + \epsilon \exp(\lambda t) \cos(k_n y)$ and 
ignoring higher orders of $\epsilon$, yields the following eigenvalues $\lambda$
(or eigenfrequencies $\omega$):
  \begin{eqnarray}
    \lambda= i \omega= \sqrt{ \frac{1}{3}\mu k_n^2 - \frac{1}{2}\Omega^2},
    \label{new1}
  \end{eqnarray}
where $k_n=n \pi/L_y$ and $L_y$ is the length of the transverse
direction (extending from $-L_y$ to $L_y$).
Importantly, this is a prediction suggesting the presence (for
large chemical potential $\mu$) of a large number of unstable eigendirections
whose growth rate is explicitly given by Eq.~(\ref{new1}). Note that in the 
large chemical potential limit, $\lambda$ 
grows proportionally to $\sqrt{\mu}$.
Hence, we obtain both explicit analytical predictions, such as
Eq.~(\ref{new1}), and the simpler (in that they reduce the dimensionality
from 2D to 1D for the evolution of the soliton filament)
effective PDE model (\ref{1D_PDE_DS})
that can be compared to the full numerical computations.
\end{itemize}

\subsection{Two-Component Case}

We now turn to the case of the dark-bright (DB) soliton stripes, which are 
two-component structures that can be viewed as ``filled'' dark soliton stripes.
DB solitons in quasi-1D BECs, first predicted theoretically in Ref.~\cite{DB} 
and then studied 
in a series of experiments
(in two- and recently generalized in three-components)~\cite{hamburg,pe1,pe2,pe3,azu,pe4,pe5,spinor}, 
feature a rather intuitive physical premise:
the dark solitons operate as an effective potential well, trapping a
bright soliton in the second component, even though this latter structure 
is not possible (by itself, i.e., in a single-component setting) for 
a self-defocusing nonlinearity~\cite{pgkdjf}.

In the 1D case of the so-called Manakov model of equal interaction 
coefficients (a very realistic case in settings such as hyperfine 
states of $^{87}$Rb~\cite{DB}), the equations for the components
$u$ and $v$, 
confined respectively by the potentials $V_d$ and $V_b$, read:
\begin{eqnarray}
  i u_t &=&  -\frac{1}{2} u_{xx} + \left[V_d + |u|^2 + |v|^2 - \mu_d \right] u,
\notag
  \\
  i v_t &=& -\frac{1}{2} v_{xx} + \left[V_b + |u|^2 + |v|^2 - \mu_b \right] v.
  \label{dbeom}
\end{eqnarray}
In this case, in the absence of external potentials, $V_d=V_b=0$, the solution for 
the DB soliton is of the form:
\begin{eqnarray}
  u &=& \sqrt{\mu_d}\, \left[\cos(\alpha) \tanh(\nu (x-{\xi})) + i \sin(\alpha)\right],
\label{eq:DB1}
\\[1.0ex]
\label{eq:DB2}
  v &=& \sqrt{{N_b \nu}/{2}}\, {\rm sech}(\nu (x-{\xi})) e^{-i \mu_b t}
    e^{i \dot{\xi} x},
\end{eqnarray}
where suitable algebraic conditions connect the soliton parameters such
as the chemical potentials $\mu_d$ and $\mu_b$,
the speed related parameter $\alpha$,
the DB soliton center position ${\xi}$ and the inverse width $\nu$,
and $N_b$, the norm of the solution (corresponding to number of particles 
in the bright component) in the $v$-component~\cite{DB}.

\begin{widetext}
In 1D, the DB free energy can then be approximated as~\cite{DB}:
\begin{eqnarray}
  G_{\rm DB,1D}=\frac{4}{3} {\cal A}^3  -2 \dot{{\xi}}^2 {\cal A} +
  N_b \left( V_b-\frac{1}{2}{V_d} \right),
\notag
\end{eqnarray}
where ${\cal A}\!=\!{\cal A}(x)\!=\!(\mu_d + N_b^2/16 - V_d(x))^{1/2} $.
Similarly to the case of the dark soliton stripe, let us now consider a DB 
soliton filament described by its position $\xi(y,t)$.
Hence, assuming  $u=u(x-{\xi}(y,t))$ and $v=v(x-{\xi}(y,t))$ and accounting
for the transverse contribution to the energy,  
$G_y = \frac{1}{2} \int \left(|u_y|^2 + |v_y|^2 \right)dx$, yields the 2D free energy:
\begin{eqnarray}
  G_{\rm DB,2D}= \int  G_{\rm DB,1D} + {{\xi}}_y^2  \left(\frac{2}{3}{\cal A}^{3} -
  \frac{1}{8}N_b^2 {\cal A} +  \frac{1}{48}N_b^3 - {{\xi}}_t^2
  \frac{8 \mu_d + N_b^2 - 8 V_d}{8 \cal A}   \right)\, dy,
  \label{db6}
\end{eqnarray}
where now ${\cal A}$ and the potential terms are evaluated at ${\xi}={\xi}(y,t)$.
The resulting equation of motion for the DB filament with longitudinal profile
given by Eqs.~(\ref{eq:DB1}) and (\ref{eq:DB2}) is obtained from $dG_{\rm DB,2D}/dt=0$
by integrating along the $x$-direction.
The resulting effective 1D PDE for $\xi(y,t)$ is particularly lengthy
and has the following 
form:
\begin{eqnarray}
  &-& 2 {\cal A}^{1/2} V_d'+ N_b (V_b'-\frac{V_d'}{2}) -4 {{\xi}}_{tt} {\cal A}^{1/2}
  + {{\xi}}_{t}^2 {\cal A}^{-1/2} V_d' 
  \nonumber
\\
  &+& \left(-{\cal A}^{1/2} V_d' - {{\xi}}_{tt} \frac{{\cal A}^{-1/2}}{4}
  (8 (\mu_d-V_d) + N_b^2) + {{\xi}}_{t}^2 \frac{V_d'}{{\cal A}^{1/2}}
  - {{\xi}}_{t}^2 (8 (\mu_d-V_d) + N_b^2) \frac{{\cal A}^{3/2}}{16} V_d'
  + \frac{N_b^2}{16} {\cal A}^{-1/2}V_d'\right) {{\xi}}_y^2
    \nonumber
\\
  &-& 2 {{\xi}}_{yy} \left[ \frac{2}{3} {\cal A}^{3/2} + \frac{{N_b}^3}{48}
  -{{\xi}}_t^2 (8 (\mu-V_d)+N_b^2) \frac{{\cal A}^{-1/2}}{8}
  -\frac{N_b^2}{8} {\cal A}^{1/2}  \right]
\nonumber
\\
&-& 2 {{\xi}}_y \left[ -V_d' {\cal A}^{1/2}  {{\xi}}_y -2 {{\xi}}_t
  {{\xi}}_{ty} (8 (\mu-V_d)+N_b^2) \frac{{\cal A}^{-1/2}}{8}+ {{\xi}}_t^2
  V_d' {{\xi}}_y {\cal A}^{-1/2} \right]
\nonumber
\\
&-& 2 {{\xi}}_y \left[ -\frac{{{\xi}}_t^2}{16}
  (8 (\mu-V_d)+N_b^2) {\cal A}^{-3/2} V_d' {{\xi}}_y + \frac{N_b^2}{16}
  {\cal A}^{-1/2} V_d' {{\xi}}_y \right]=0.
\label{dbeqn}
\end{eqnarray}
\end{widetext}
Nonetheless, linearizing around the fixed point $X_0$ which bears
no $y$-dependence, using ${\xi}=X_0 + \epsilon \cos(k_n y) X_1(t)$,
we obtain the following dynamical equation for perturbations $X_1$
around the stationary (straight) filament:
\begin{eqnarray}
  {X_1}_{tt}= -\omega_n^2 X_1, 
\notag
\end{eqnarray}
with (squared) eigenfrequencies
\begin{eqnarray}
  \omega_n^2&=&\frac{1}{2} V_d'' - \frac{N_b}{4 {\cal A}_0}
    \left(V_b''-\frac{1}{2}V_d''\right) 
\notag
\\[1.0ex]
\notag
    &&-  k_n^2 \left(
    \frac{1}{3}{\cal A}_0^2 + \frac{1}{96} \frac{N_b^3}{ {\cal A}_0 }
    - \frac{1}{16} N_b^2\right),
\end{eqnarray}
where now ${\cal A}_0=\left.{\cal A}\right|_{{\xi}=X_0}$, and all 
potentials
(and their derivatives) are evaluated at $X_0$.
For the experimentally relevant 
case of a parabolic trap $V_b=V_d=\frac{1}{2} \Omega^2 x^2$
~\cite{pgkdjf}, we have
$V_d(X_0)=V_b(X_0)=V_d'(X_0)=V_b'(X_0)=0$, $V_d''(X_0)=V_b''(X_0)=\Omega^2$,
and ${\cal A}_0=(\mu_d+N_b^2/16)^{1/2}$, leading to:
\begin{equation}
  \omega_n^2 = \frac{1}{2}\Omega^2
  - \frac{N_b}{8{\cal A}_0} \Omega^2
  - \frac{1}{3}\mu_d k_n^2
  - \left( \frac{N_b}{4{\cal A}_0}-1 \right) \frac{N_b^2\, k_n^2}{24} .
    \label{db10}
\end{equation}

We can now make the following relevant observations regarding the 
eigenfrequencies given in Eq.~(\ref{db10}):

\begin{itemize}
\item[(i)]
  The first term represents the oscillation frequency of 
  the 1D dark soliton in a trap~\cite{busch,konotop_pit};
  the second term constitutes the correction to this frequency
  in the DB soliton stemming from the bright component
  (still in 1D)~\cite{DB}.

\item[(ii)]
  The third term is the transverse undulation frequency contribution
  from a flat background (in the transverse direction). Together,
  the first and third term combine to yield the result of
  Eq.~(\ref{new1}) for the undulation in a 1D trap of the 2D
  dark soliton stripe.

\item[(iii)]
  Finally, the fourth term corresponds to the contribution to the 2D
  transverse undulation stemming from the bright soliton.
\end{itemize}

An appealing feature of this step-by-step approach is that one not only
obtains an expression for the spectral mode eigenfrequencies, but also
an intuitive sense on the nature and origin of {\em each} contribution.

Having explored both the nonlinear (fully dynamical)
and the linear (spectral) setup of such
a multi-component soliton filament, it is natural to examine how these
conclusions fare against the full numerical computations
of the original 2D model of Eq.~(\ref{extr}).

\section{Numerical Methods and Findings}

\subsection{General setup and methodology}

In our numerical simulations, we consider the full 2D dimensionless GP equations 
(\ref{extr}) and (\ref{dbeom}) for the one- and two-component cases, respectively.
We consider a trapping potential acting only 
along the $x$-direction, namely:
\begin{equation}
\label{eq:VMT}
V(x,y)=\frac{1}{2} \Omega^2 x^2,
\end{equation}
and we consider periodic boundary conditions along the $y$-direction.
As for the trap strength, we use 
---without loss of qualitative generality of our results--- 
$\Omega=1$ for all of the following numerical computations.

Our numerical simulations consist of the following serial steps: we first 
solve for stationary states and compute their linear stability spectrum,
and, finally, we explore their dynamics. Because the system 
has $y$-translational symmetry (due to the form of the potential 
and its associated steady states), we solve 
the stationary states only along the $x$-direction to obtain the
cross section of the sought-for 2D steady states. 

Furthermore, to render the 2D stability computations more efficient,
we use the fact that our solutions are $y$-independent in order to extract the
linear stability eigenvalues as a collection of 1D eigenvalue problems
using basis expansions, also called the partial wave method. 
This technique is summarized in Refs.~\cite{wenlong,wenlong2} for one- and 
two-component radially symmetric BECs.
The method can be 
straightforwardly tailored in a similar manner to our setting by replacing
the angular direction $\theta$ with $y$ (and expressing the Laplacian
in rectangular coordinates rather than polar). 
Since the methods 
are fairly similar, we refer the interested reader to Refs.~\cite{wenlong,wenlong2}
for more details.
Nonetheless, we briefly mention here that the method computes eigenvalues for 
each $y$-mode separately (in our case $k_n$ or $n$, and eigenvalues of $k_n$ 
and $-k_n$ are complex conjugates) and the full 2D spectrum is simply the union
of all the individual 1D spectra.

\begin{figure}[htb]
\begin{center}
\includegraphics[width=\columnwidth]{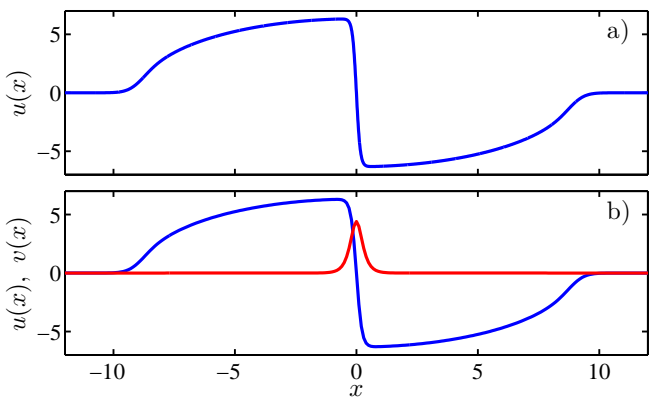}
\caption{(Color online) 
Cross sections ($y={\rm const.}$) along the $x$-direction of representative waves 
corresponding to (a) the dark soliton,
for $\mu=40$, and (b) the dark-bright soliton, for $\mu_d=40$ 
and $\mu_b=29.682$ [the dark (bright) component is depicted in blue (red)]; in 
both cases $\Omega=1$. 
Note that these 2D stationary states are 
homogeneous in the $y$-direction as the potential (\ref{eq:VMT}) is
only $x$-dependent. 
}
\label{fig:States}
\end{center}
\end{figure}

In our computations, we use the domain $x\in[-16,16]$ which is sufficiently long 
to support the background cloud carrying the dark and dark-bright solutions, 
and we use chemical potentials
up to $\mu=80$. 
We have checked that the domain size along the $x$-direction
(provided it is large enough to support the background cloud) 
does not affect the 
numerical results hereby presented.
In Fig.~\ref{fig:States} we depict a typical example (in the case
of large chemical potentials) for
the dark and dark-bright soliton states.

In what follows, we span the spectra of the original NLS model
using the lowest $n=0, 1, 2, ..., 10$ modes, as in Ref.~\cite{ai}. 
In our simulations, a typical lattice spacing for the finite difference method 
is $\delta x=0.001$, and in certain cases, a small spacing as low as 
$\Delta x=0.0002$ was required to achieve spectrum convergence at high densities. 
The full PDE dynamics were performed using a standard second order finite differencing
in space combined with a forward fourth-order Runge-Kutta in time.

\begin{figure}[t]
\begin{center}
\includegraphics[width=8.5cm]{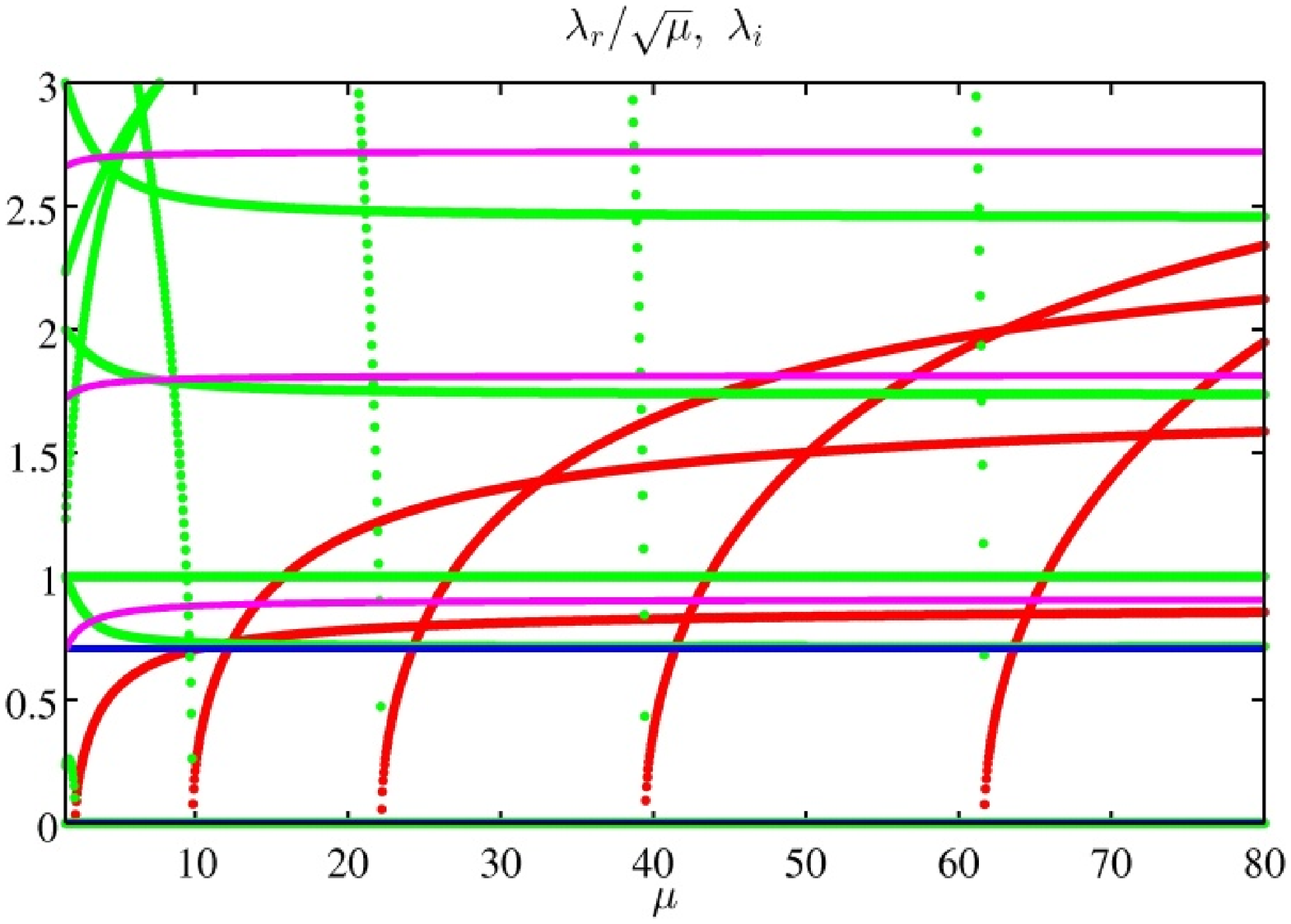}
\\[1.0ex]
\includegraphics[width=8.5cm]{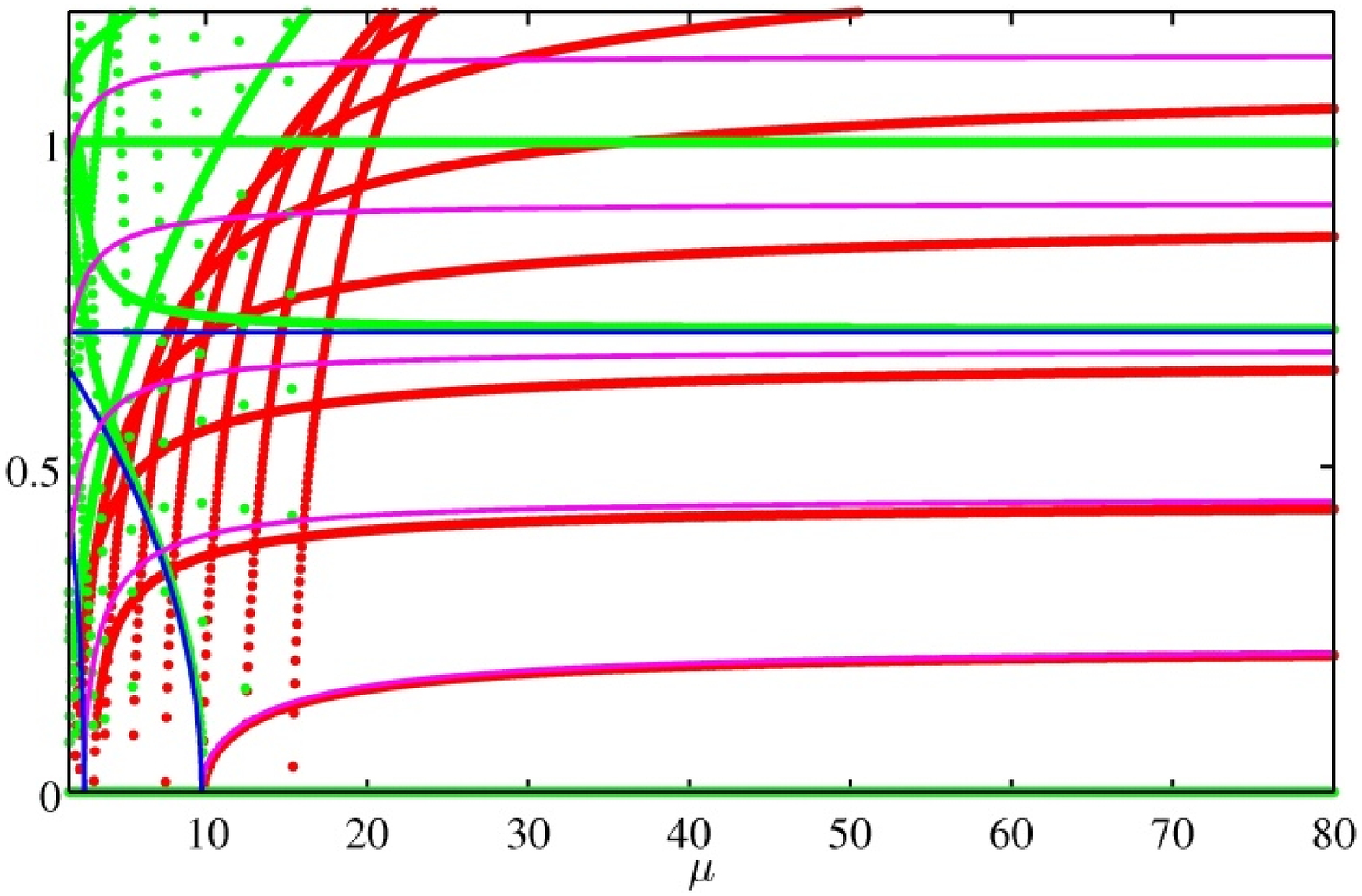}
\caption{(Color online) 
Comparison between the dark soliton stripe stability spectra for the full NLS model 
(\ref{dark1}) and the analytical prediction (\ref{new1}) for its reduced AI variant.
Depicted are the stability eigenvalues 
$\lambda=\lambda_r + i\,\lambda_i$ as a function of the chemical 
potential $\mu$. The numerical domain is $(x,y)\in[-L_x,L_x]\times[-L_y,L_y]$
with $L_x=16$ and $L_y=2$ (top panel) and $L_y=8$ (bottom panel). 
The real part $\lambda_r$ of the eigenvalue 
is scaled by $\sqrt{\mu}$.
Red and green dots correspond to the real and imaginary parts of the spectrum
from the full NLS model while pink and blue lines correspond to the real
and imaginary parts for the effective AI model.
}
\label{DS1}
\end{center}
\end{figure}

\subsection{NLS and AI spectra}

Now that we are equipped with the reduced AI PDEs (\ref{1D_PDE_DS}) and
(\ref{dbeqn}) for
the dark and the DB solitons for one- and 
two-components, respectively, let us corroborate the validity of this reduction
approach at the level of the associated spectra for stationary states.
Thus, we 
numerically compute the spectra for the dark and DB solitons
as the chemical potential $\mu$ is varied starting from the linear limit.
The dark soliton emerges from the linear limit at $\mu=3/2$ as it is the 
first excited state of the (1D) 
quantum harmonic oscillator. Similarly,
the DB soliton emerges from the linear limit at $\mu_d=3/2$ and $\mu_b=1/2$
corresponding to coupling the first excited state and the ground state of the 
quantum harmonic oscillator. 
We follow the dark soliton steady state configuration and its corresponding 
spectrum using continuation starting from the linear 
limit ($\mu=3/2$) up to $\mu=80$.

The spectra for both the NLS model (\ref{dark1}) and our analytical
prediction (\ref{new1}) for the effective AI reduction are depicted in Fig.~\ref{DS1},
for two values of the transverse length $L_y$:
the top panel corresponds to a relatively small $L_y=2$, while
the bottom panel corresponds to $L_y=8$.
As expected, the stability properties of the dark soliton stripe strongly depend
on the domain's transverse length $L_y$. In particular, a larger number of
instabilities are observed for larger values of $L_y$ since larger domains can 
support instabilities with shorter wavenumbers.
However, more importantly, we observe that the NLS and AI spectra agree 
reasonably well, with better agreement for larger chemical potential $\mu$.
Moreover, the lower frequency (and/or growth rate) modes converge
well for smaller chemical potentials, while the larger frequency
(and growth rate) modes are progressively better for higher chemical
potentials.

\begin{figure}[htb]
\begin{center}
\includegraphics[width=8.5cm]{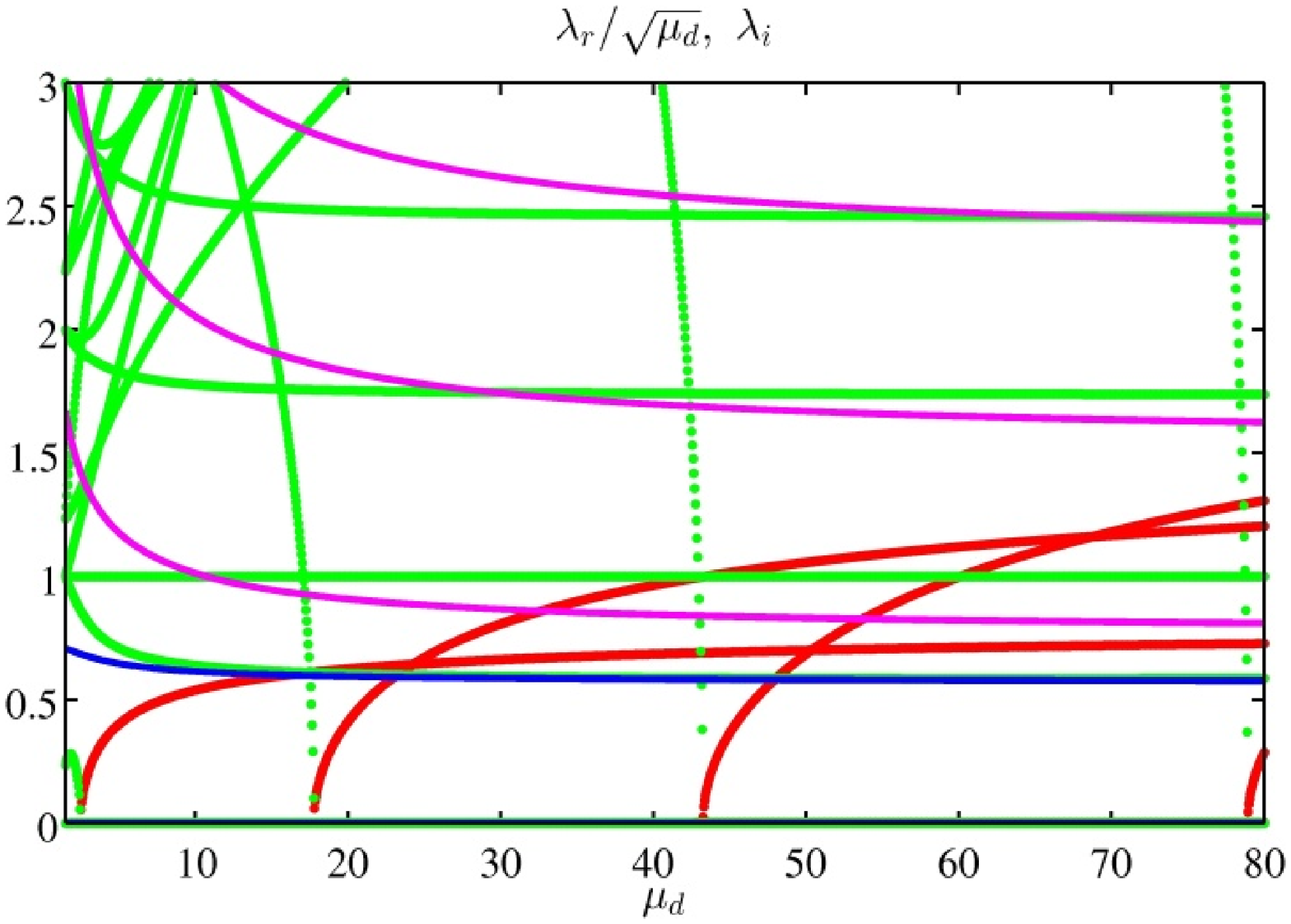}
\\[1.0ex]
\includegraphics[width=8.5cm]{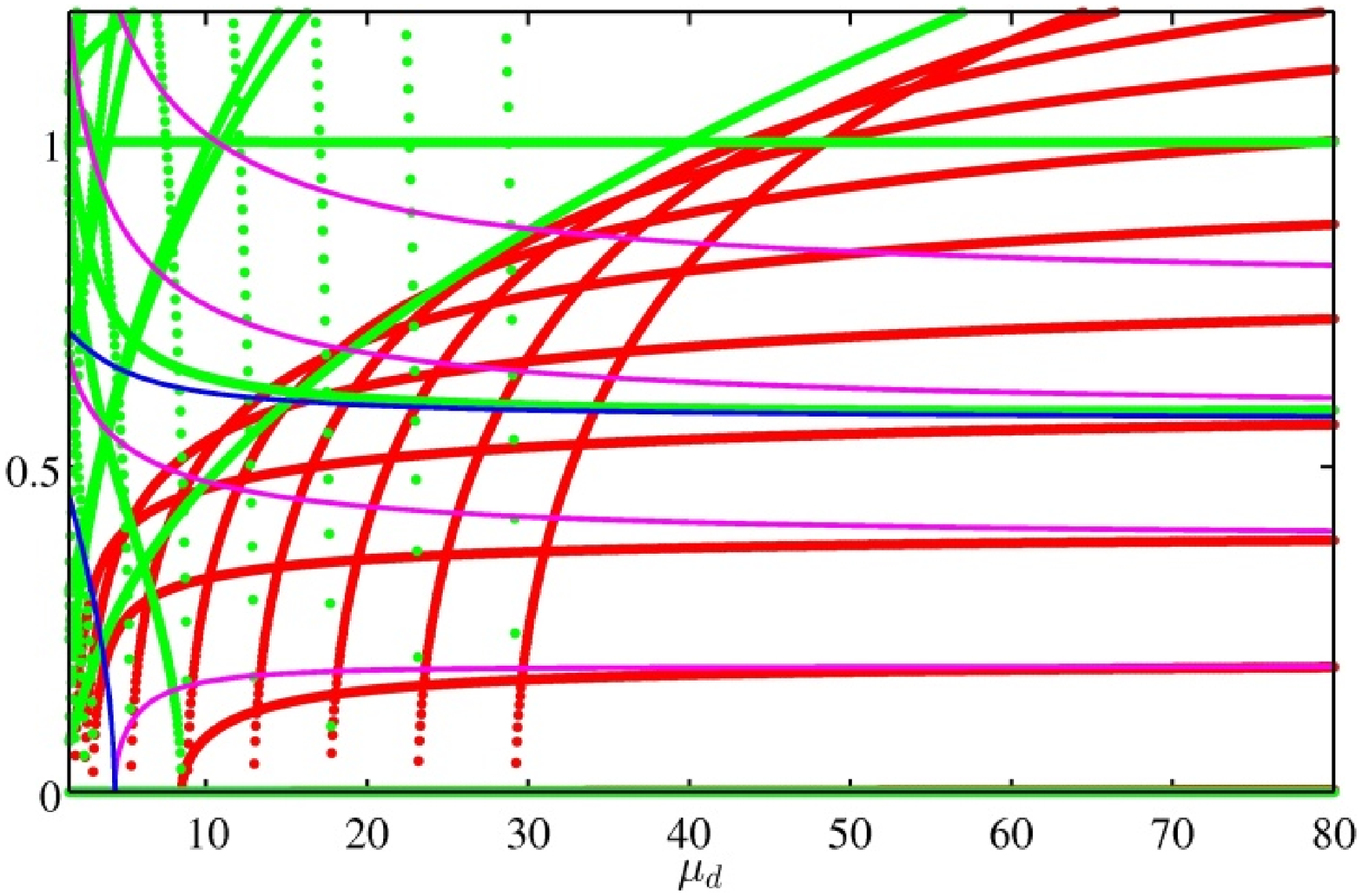}
\caption{(Color online) 
Comparison between the DB soliton stripe stability spectra for the full NLS model 
(\ref{dbeom}) and the analytical prediction (\ref{db10}) for its reduced AI variant.
Same layout and parameters as in Fig.~\ref{DS1}.
Here, Re($\lambda$) is scaled by $\sqrt{\mu_d}$ and
the $x$-axis corresponds to a $\mu_d$ and $\mu_b$ combination given by 
the linear ``trajectory'' in $(\mu_d,\mu_b)$ parameter space starting from 
the linear limit $(\mu_d,\mu_b)=(3/2,1/2)$ to the final value
$(\mu_d,\mu_b)=(80,60)$. 
}
\label{DBS1}
\end{center}
\end{figure}

Figure~\ref{DBS1} depicts a similar scenario to the dark soliton stripe
presented in Fig.~\ref{DS1}, but now for the DB soliton stripe. In this case, 
we start from the linear limit $(\mu_d,\mu_b)=(3/2,1/2)$ and progress
with a linear ``trajectory'' in the $(\mu_d,\mu_b)$ parameter space until
reaching $(\mu_d,\mu_b)=(80,60)$. As before, there is a very good agreement 
between the full NLS spectrum and the analytical prediction (\ref{db10}) computed 
from the AI reduction.

It is interesting to note that, despite the strong instabilities present at high 
densities, both dark and DB soliton stripes can be stable sufficiently close to 
the linear limit. This stabilization is due to the finite size of the domain
in the $y$-direction, where small enough wavenumbers will not be able to fit
in the domain. For instance, when $L_y=2$, the dark soliton stripe does not
acquire an unstable eigenvalue until reaching $\mu \simeq 2.10$. In fact, we have
checked numerically that full (2D) time integration of the stationary 
dark soliton stripe for $\mu=2$ is indeed stable for long times (results not shown here).
Similarly, the spectrum for the DB soliton stripe suggests that this configuration
[for the choice of $(\mu_d,\mu_b)$ parameters described above] is stable 
for $\mu_d \lesssim 2.45$ [along the aforementioned $(\mu_d,\mu_b)$ parameter 
trajectory]. We have also verified, by direct integration, that the DB soliton 
stripe for $(\mu_d,\mu_b)=(2.4,1.1822)$ is indeed stable for long times
(results not shown here). 
For both simulations we added to the exact stationary stripe states a relatively 
small random perturbation (on the order of $10^{-8}$), and no visible instability
growth was observed for times up to $t=1000$.

\begin{figure}[htb]
\begin{center}
\includegraphics[width=\columnwidth]{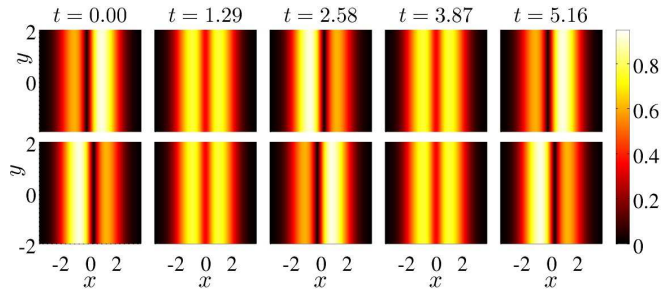}
\caption{(Color online) 
Stable dark-dark soliton stripe oscillations
in a two-component NLS.
The transverse domain length $L_y=2$ is small enough to arrest any potential
instabilities for the chemical potentials $(\mu_d,\mu_b)=(2.4,1.1822)$.
The period of the breathing pattern is 
$T=2\pi/(\mu_d-\mu_b)=2\pi/(2.4-1.1822) \approx 5.1595$ (cf.~Ref.~\cite{wenlong3} 
for details), which matches extremely well the observed period of the dynamics. 
The top (bottom) panels depict snapshots of the density for the first 
(second) components at times, from left to right, $t=0, T/4, T/2, 3T/4, T$, 
respectively. 
The two dark solitons start from opposite sides of the trap, move together 
and pass through each other, reaching the other sides, and oscillate back. 
}
\label{DBST}
\end{center}
\end{figure}

The stability for small enough values of the chemical potential (and/or small
enough domain lengths $L_y$) can be used to
stabilize additional solutions.
For instance, it is possible to stabilize two-component breathing dark-dark 
soliton stripes, resulting 
from a SO(2) rotation of DB soliton stripes, similarly to the quasi-1D case 
\cite{pe4,pe5,wenlong3}.
These solutions are based on two coupled dark solitons with different chemical 
potentials, $\mu_d \ne \mu_b$. 
An example of such a 
stable breathing dark-dark soliton stripe 
is depicted in Fig.~\ref{DBST}, with the top and bottom panels showing each of the 
two components. It is observed that the two dark soliton stripes 
start from different sides of the trap, pass through each other, and oscillate back 
to restart the cycle. We have checked that, indeed, this oscillating pattern is stable
and that the 
oscillation period $T$ is indeed determined by the 
chemical potential imbalance $\mu_d-\mu_b$, namely 
$T=2\pi/(\mu_d-\mu_b)$ (cf.~Ref.~\cite{wenlong3} for details on 
the derivation of this result).

\begin{figure*}[htb]
\begin{center}
\includegraphics[width=2\columnwidth]{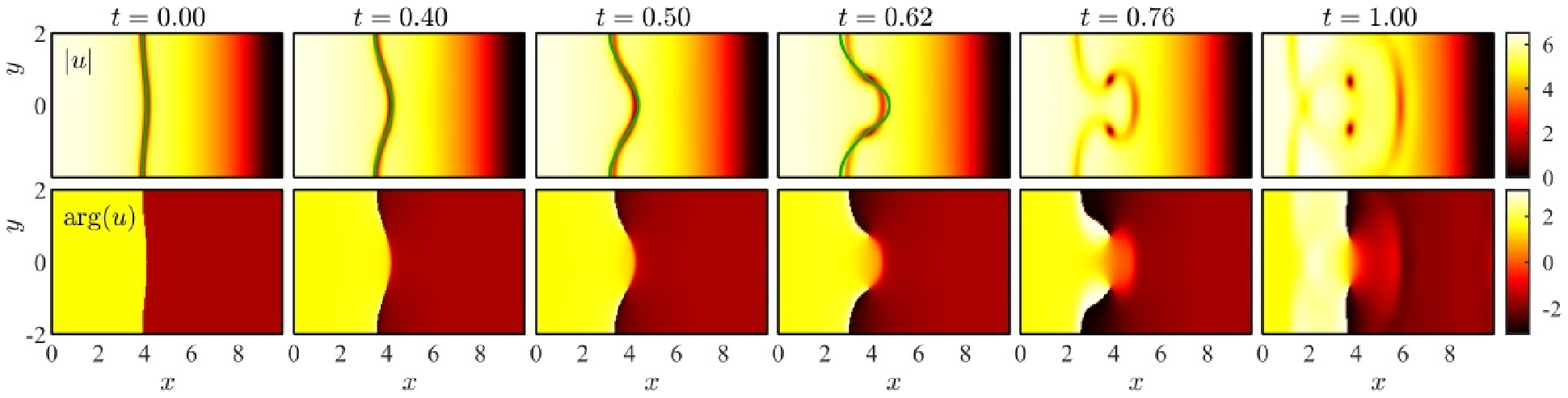}
\\[1.0ex]
\includegraphics[width=2\columnwidth]{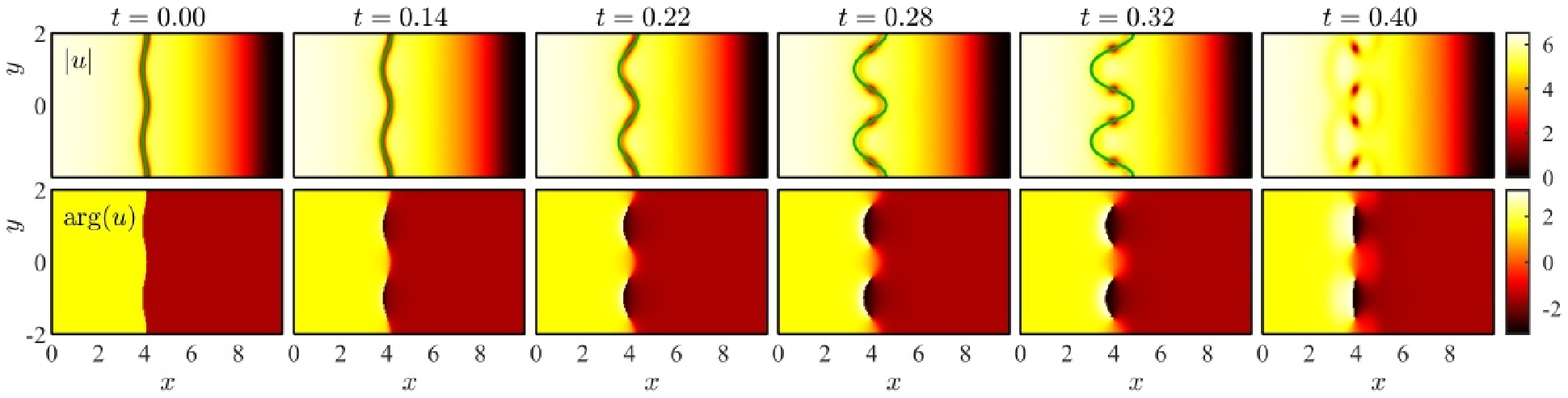}
\caption{(Color online) 
Dynamical destabilization of the dark soliton stripe
corresponding to the full NLS (\ref{dark1}) [see 
background colormap] and the AI reduction (\ref{1D_PDE_DS}) 
[see (green) curves in the corresponding top row in each
set of panels].
The corresponding systems are initialized with a dark
soliton stripe at $x(y) = x_0 + A\cos(n \pi y/L_y)$ with
$x_0=4$, $A=0.1$, with $\mu=40$, and $n=1$ (top set of panels) 
and $n=2$ (bottom set of panels).
Within each set of panels the top and bottom row correspond
to the magnitude ($|u(x,t)|$) and phase of the field at the
indicated times.
We note that, for better comparison between frames, the phase has been 
rotated so that, for all times, the phase at the origin is fixed to $\pi/2$.
We also note that, for better visibility of the destabilization features, 
the panels only depict the domain for $x\geq 0$ (the $x<0$ region
has trivial dynamics as there is no stripe in there).
See supplemental movies {\tt DS1\symbol{95}movie} and  {\tt DS2\symbol{95}movie}.
}
\label{fig:PDE_AI_DS}
\end{center}
\end{figure*}

\subsection{NLS and AI dynamics}
\label{dynamics}

In this last section, we compare the evolutionary dynamics
for dark and DB soliton 
stripes obtained through the AI reduction and the original NLS model. 

First, we compare the dynamics of the dark soliton stripe, as described by 
the AI reduction and obtained by the original NLS model. 
For all the comparisons presented below we chose a relatively large
chemical potential $\mu=40$ for the dark component; recall that for this relatively
large value of the chemical potential, we concluded that there is a good match
between the corresponding spectra of  the AI reduction and of the 
NLS model.
Furthermore, in order to keep at bay the amount of instabilities that
can be present in the system, we use a relatively small transverse length
of $L_y=2$ for which only a limited number of instabilities are present
(see the previous section for details).

To initialize the system we consider a dark soliton stripe initially displaced in 
the $x$-direction by $x_0$ and perturbed in the (transverse) $y$-direction by $n$ 
harmonic undulations of amplitude $A$. To be more specific, this amounts 
to a filament with initial position given by $x(y) = x_0 + A\cos(n \pi y/L_y)$, and with
zero initial velocity; 
in what follows we use $x_0=4$, $A=0.1$, and $n=1,2$ for all of our numerics.
This initial perturbation is intended to seed a specific destabilization
eigendirection for best comparison between the AI and NLS models.
Choosing random initial perturbations along the $y$-direction
results in similar destabilizations along the {\em most unstable}
eigendirection, but the precise timing and the location (along the
$y$-axis) of the unstable mode obviously depend on each realization; 
thus, a match between the AI and NLS models is less straightforward to achieve.
Since the initial condition does not correspond to a steady state,
and since we do not have access to the exact left-to-right oscillatory 
solution of a dark, or DB, soliton, we initialize the NLS model
with the corresponding displaced (to $x_0$) exact solution (found
in the absence of external potential, $V(x)=0$) with a local chemical 
potential adjusted to $\mu-V(x-x_0)$, as per the adiabatic invariant
approach.

In general, we expect the evolution of the stripes to adhere to two
principal features: 
\begin{itemize}
\item[(i)] the left-to-right oscillations ---with frequencies $\Omega/\sqrt{2}$ for the 
dark soliton and the corresponding adjusted frequency (\ref{db10}) due to the 
presence of the bright soliton components for the DB soliton---
and 
\item[(ii)] the destabilization of the stripe through the perturbed $n$-th
  undulation mode
(if it is indeed unstable).
\end{itemize}
The former trait, for our choice of $\Omega=1$, leads
to a left-to-right
oscillation period of around $2\pi$. 
In contrast, note that the instabilities ---see spectra of the previous section--- 
have typical values of order one when {\em divided} by $\sqrt{\mu}$.
In fact, the instabilities for large $\mu$ scale precisely as $\sqrt{\mu}$
and, thus, for the chosen relatively large chemical value of $\mu=40$,
the instabilities will grow proportional to $e^{\sqrt{\mu}t} \sim e^{6.3t}$.
Therefore, the growth of instabilities will be typically much faster
than the left-to-right oscillations and thus the latter oscillations
will not be typically observable within the time range of our simulations
focusing on the growth of the instabilities.

\begin{figure*}[htb]
\begin{center}
\includegraphics[width=2\columnwidth]{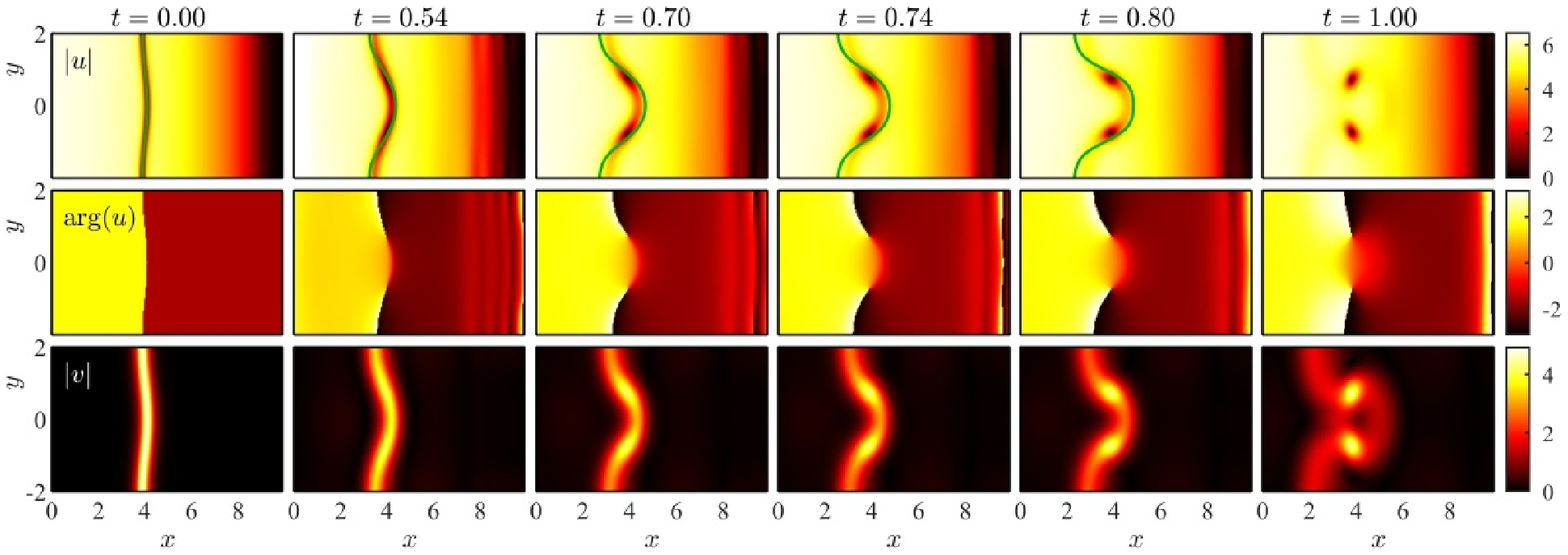}
\\[1.0ex]
\includegraphics[width=2\columnwidth]{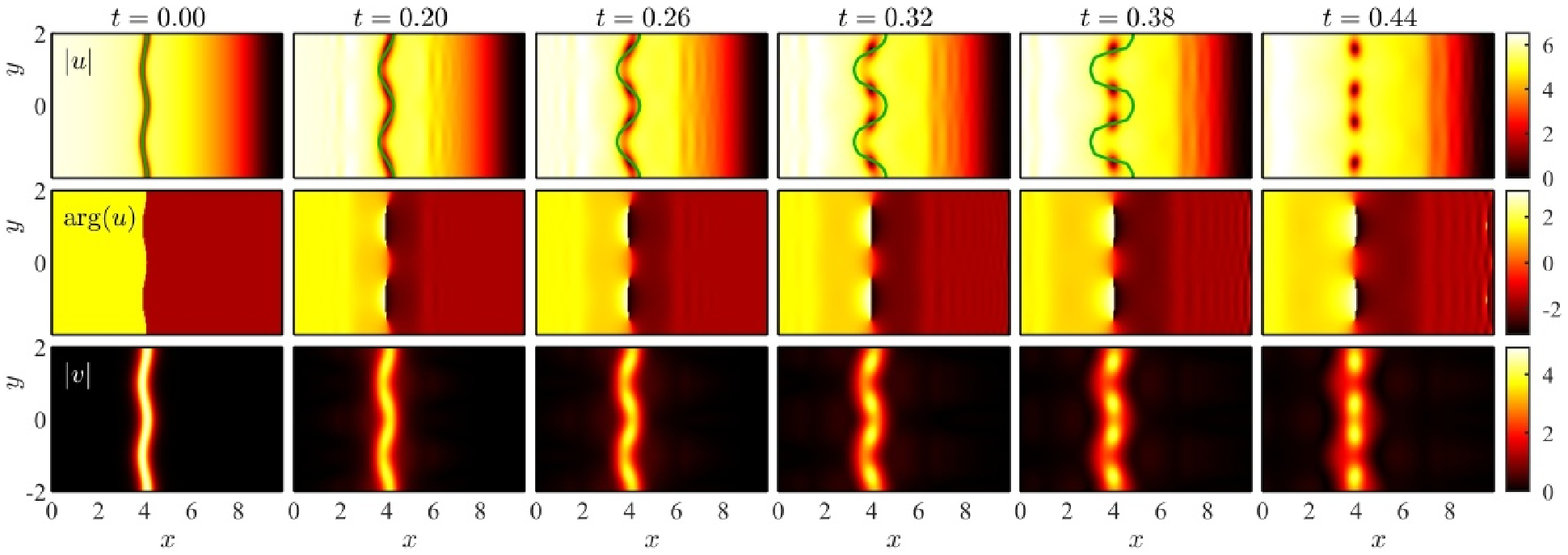}
\caption{(Color online) 
Dynamical destabilization of the DB soliton stripe
corresponding to the full NLS (\ref{dbeom}) [see
background colormap] and the AI reduction (\ref{dbeqn})
[see (green) curves in the corresponding top row in each
set of panels].
The perturbations to the initial DB soliton stripe and layout of the 
figure are the same as in Fig.~\ref{fig:PDE_AI_DS} with the
addition of a third row of panels depicting the magnitude
of the second field ($|v(x,t)|$). The values of the chemical
potentials are 
$\mu_d=40$ and $\mu_b=29.6815$.
See supplemental movies {\tt DB1\symbol{95}movie} and  {\tt DB2\symbol{95}movie}.
}
\label{fig:PDE_AI_DB}
\end{center}
\end{figure*}

Figure~\ref{fig:PDE_AI_DS} depicts two examples for the destabilization
of the dark soliton stripe through the $n=1$ (top set of panels)
and $n=2$ (bottom set of panels) modes.
As expected, the left-to-right oscillation of the dark soliton
stripe is barely visible while the stripe suffers a strong instability
along the $n=1$ and $n=2$ modes. This instability is responsible for
spatial undulations that
the dark 
soliton filament develops, the so-called
{\em snaking}, along the $y$-direction.
As the snaking intensifies, the filament breaks up into pairs of vortices
(see dark spots in the field's modulus and the $2\pi$ phase jump
singularities in the field's phase). In fact, $n$ pairs of vortices
are created when perturbing with the $n$-th mode.
More importantly, the figure shows that the reduced dynamical AI model
(\ref{1D_PDE_DS}) is able to qualitatively and quantitatively describe
the full NLS  evolution of Eq.~(\ref{dark1})
before the filament breakup
into vortex pairs.
Note that the AI approach displays a slightly faster instability
growth rate when
compared to the original NLS dynamics. This is straightforwardly understandable
as the AI spectra predicts slightly larger real parts for the eigenvalues when 
compared to the original NLS dynamics (see, for instance, the
top panel in Fig.~\ref{DS1}).
Also notice that the AI results are not shown past the time when the
filament starts to develop the vortex pairs. At that point, the
AI PDE solution develops singularities (vertical slope) and its numerical
evolution
breaks down. This is of course natural as, by construction, our
AI dynamics:
\begin{itemize}
\item[(i)] does not allow for bends of the filament leading to 
multi-valuedness of the filament's location $x(y)$, and 
\item[(ii)] as the
original NLS filament starts breaking up into vortex pairs, the 
assumption that the solution remains as a longitudinal dark soliton
filament is clearly violated.
\end{itemize}
Nonetheless, it is remarkable that the lower dimensional AI reduction
is able to qualitatively, and, where appropriate, even
quantitatively, capture the soliton filament
dynamics before its breakup into vortex pairs.

In Fig.~\ref{fig:PDE_AI_DB} we present results similar 
to the ones presented in Fig.~\ref{fig:PDE_AI_DS}, but for the DB soliton stripe. 
The conclusions stated above also apply to this more complex case, where 
our AI approach is 
able to capture the snaking of the DB soliton filament before
its breakup into vortex pairs in the dark component filled by bright
cores in the other component.
The latter vortex-bright single and pair structures have also
been previously examined; see, e.g., Refs.~\cite{law,pola} and references therein.
It is also interesting to note that the instability rates
for the DB soliton stripe are somewhat reduced
when compared to the pure
dark soliton stripe. Therefore, the observed time for the filament to 
experience breakup into vortex pairs is correspondingly increased for the
DB stripe when compared to its pure dark stripe counterpart.
In fact, the quantitative specifics of the instability reduction depend on the
mass of the bright component which serves as an effective repulsive potential taming 
the destabilization of the dark component, in agreement with previous results~\cite{us,braz}. 
In our specific numerical experiments the pure dark soliton stripe starts the
vortex pair breakup around $t\sim 0.64$ and $t\sim 0.28$ for the $n=1$ and $n=2$
modes respectively. In contrast, the DB soliton stripe does not suffer
the vortex pair breakup until $t\sim 0.84$ and $t\sim 0.34$ 
for the $n=1$ and $n=2$ modes respectively.

\section{Conclusions \& Future Work}

In the present work, we have examined the existence, stability and
dynamical properties of the evolution of soliton 
filaments ---i.e., quasi-one-dimensional structures--- 
embedded in higher-dimensional settings (in particular, two-dimensional ones 
in the present context).
We did so both for the simpler case of the single-component
dark soliton stripe, as well as for the technically more involved
case of the dark-bright soliton in the two-component setting.
The employed adiabatic
invariant approach enables the formulation of a partial differential
equation at reduced dimensionality, i.e., going from a two-dimensional
field $u=u(x,y,t)$ to a one-dimensional characterization 
for the evolution of the filament position $\xi=\xi(y,t)$. 
Additionally, the nature
of the formulation endows it with a Hamiltonian structure.

A fundamental advantage of the formulation is that perturbations
around the steady-state rectilinear stripe can be considered in an analytical form, 
and explicit expressions for the linearization eigenfrequencies
tracking the ``undulations'' of the filamentary structure can
be identified. 
These modes are responsible for the
transverse (snaking) instability, leading to the breakup of the structure,
hence this approach enables insights into the relevant modes 
and their growth rates. Parametric dependences (e.g., on the
number of atoms of the bright component) can also be identified within the
model. Moreover, through numerics, the approach allows
for a lower-dimensional (i.e., quasi-one-dimensional in the
settings considered herein) visualization
of the system dynamics that remains faithful to the full (higher-dimensional)
PDE dynamics until the vicinity of the relevant breakup time
towards vortices (or vortex-bright solitons
in the multi-component case) as a result of the transverse instability.

It is worthwhile to consider whether the success of the method
can be generalized to other settings. Perhaps a simple one
to state,  
although challenging to set up, is the scenario where the
rectilinear stripe is examined in the case of a two-dimensional
parabolic trap (i.e., finite trapping along both directions); 
see, e.g., Ref.~\cite{middel_pra}. There, the quasi one-dimensional
nature of the configuration is no longer present and, hence, a suitable 
amendment of the technique, to account for the finite length of the 
filament and its modification close to the boundary edges,
needs to be considered. In the context of the two-component setting,
extending the considerations presented herein to the case of
a dark-bright ring is a natural next step, allowing to expand
on the radial considerations of Ref.~\cite{jans}. Finally, a more
demanding scenario to consider, in the sense that it involves
multiple PDEs or a single PDE in a higher-dimensional set up, 
is that of the examination of vortex rings and vortex lines
in three-dimensional condensates~\cite{fetter1}. Such studies are
presently in progress and will be reported in future publications.

\acknowledgments 

W.W.~acknowledges support from the Swedish Research Council Grant 
No.~642-2013-7837 and Goran Gustafsson Foundation for Research in
Natural Sciences and Medicine.
P.G.K.~gratefully acknowledges the support of
NSF-PHY-1602994, as well as from  the Greek Diaspora
Fellowship Program. 
R.C.G.~acknowledges support from PHY-1603058.


\begin{thebibliography}{99}

\bibitem{Kivshar-LutherDavies} 
Yu. S. Kivshar and B. Luther-Davies,
Phys.\ Rep.\ {\bf 298}, 81--197 (1998).

\bibitem{djf} D. J. Frantzeskakis,
J. Phys. A: Math. Theor. {\bf 43}, 213001 (2010).

\bibitem{Pismen1999} 
L. M. Pismen, 
{\it Vortices in Nonlinear Fields} (Clarendon, UK, 1999).


\bibitem{YSKPiO}  
Yu. S. Kivshar, J. Christou, V. Tikhonenko, B. Luther-Davies and L. Pismen,
Optics Comm.\ {\bf 152} (1998) 198--206.

\bibitem{desa} A. S. Desyatnikov, L. Torner, and Yu. S. Kivshar, 
Prog. Opt. {\bf 47}, 291--391 (2005).

\bibitem{fetter1} 
A. L. Fetter and A. A. Svidzinsky,
J. of Phys.: Condensed Matter {\bf 13}, R135--R194 (2001).

\bibitem{fetter2}
A. L. Fetter,
Reviews of Modern Physics {\bf 81}, 647--691 (2009).

\bibitem{kuzne} E. A. Kuznetsov and S. K. Turitsyn,
Zh. Eksp. Teor. Fiz. {\bf 94}, 119--129 (1988) 
[Sov. Phys. JETP {\bf 67},  1583--1588 (1988)].


\bibitem{kidep} Yu. S. Kivshar and D. E. Pelinovsky,
Phys. Rep. {\bf 331}, 117--195 (2000).

\bibitem{tikh} V. Tikhonenko, J. Christou, B. Luther-Davies, 
and Yu. S. Kivshar, 
Opt. Lett. {\bf 21}, 1129--1131 (1996).

\bibitem{watching} B. P. Anderson, P. C. Haljan, C. A. Regal, D. L. Feder, 
L. A. Collins, C. W. Clark, and E. A. Cornell,
Phys. Rev. Lett. {\bf 86}, 2926--2929 (2001). 

\bibitem{smirnov} V. A. Mironov, A. I. Smirnov, and L. A. Smirnov,
  Zh. Eksp. Teor. Fiz. {\bf 139}, 55 (2011) [Sov. Phys. JETP {\bf 112}, 46 (2011)].

\bibitem{hoefer} M. A. Hoefer and B. Ilan, Phys. Rev. A {\bf 94}, 013609 (2016). 

\bibitem{ai} P. G. Kevrekidis, W. Wang, R. Carretero-Gonz{\'a}lez, and D. J. Frantzeskakis,
Phys. Rev. Lett. {\bf 118}, 244101 (2017).
  
\bibitem{us} 
M. Ma, R. Carretero-Gonz{\'a}lez, P. G. Kevrekidis, D. J. Frantzeskakis, 
and B. A. Malomed, Phys. Rev. A {\bf 82}, 023621 (2010) and references therein.


\bibitem{kivyang} 
Yu. S. Kivshar and X. Yang, Phys. Rev. E {\bf 50}, R40 (1994).

\bibitem{rings} D. Neshev, A. Dreischuh, V. Kamenov, I. Stefanov, S.
Dinev, W. Fliesser, and L. Windholz, Appl. Phys. B
{\bf 64}, 429 (1997); A. Dreischuh, D. Neshev, G. G. Paulus,
F. Grasbon, and H. Walther, Phys. Rev. E {\bf 66}, 066611
(2002).

\bibitem{rings1} T. P. Horikis and D. J. Frantzeskakis, 
Opt. Lett. {\bf 41} 583--586 (2016).


\bibitem{rings2} 
G. Theocharis, D. J. Frantzeskakis, P. G. Kevrekidis, 
B. A. Malomed, and Yu. S. Kivshar, Phys. Rev. Lett. {\bf 90}, 120403 (2003).

\bibitem{rings3} G. Theocharis, P. Schmelcher, M. K. Oberthaler, 
P. G. Kevrekidis, and D. J. Frantzeskakis, 
Phys. Rev. A {\bf 72}, 023609 (2005).

\bibitem{rings4} L. A. Toikka, J. Hietarinta, and K.-A. Suominen, 
J. Phys. A: Math. Theor. {\bf 45}, 485203 (2012).


\bibitem{carr} 
L. D. Carr and C. W. Clark, Phys. Rev. A {\bf 74}, 043613 (2006).
 
\bibitem{wenlong} 
W. Wang, P. G. Kevrekidis, R. Carretero-Gonz{\'a}lez, and D. J. Frantzeskakis,
Phys. Rev. A {\bf 93}, 023630 (2016).

\bibitem{hau} N. S. Ginsberg, J. Brand, and L. V. Hau, Phys. Rev. Lett. {\bf 94},
040403 (2005).

\bibitem{konotop_pit} V. V. Konotop and L. P. Pitaevskii,
Phys. Rev. Lett. {\bf 93}, 240403 (2004).

\bibitem{konotop_pit2} V. A. Brazhnyi, V. V. Konotop, and L. P. Pitaevskii,
 Phys. Rev. A {\bf 73}, 053601 (2006).

\bibitem{korneev} A. M. Kamchatnov and S. V. Korneev,
   Phys. Lett. A {\bf 374}, 4625 (2010).

\bibitem{rcg:BEC_book2}
P. G. Kevrekidis, D. J. Frantzeskakis, and R. Carretero-Gonz{\'a}lez,
{\it The defocusing nonlinear Schr{\"o}dinger equation: from dark solitons
and vortices to vortex rings} 
(SIAM, Philadelphia, 2015).

\bibitem{busch} Th. Busch and J. R. Anglin,
Phys. Rev. Lett. {\bf 84} 2298--2301 (2000).  

\bibitem{DB} Th. Busch and J. R. Anglin,
Phys. Rev. Lett. {\bf 87}, 010401 (2001).

\bibitem{hamburg}
C. Becker, S. Stellmer, P. Soltan-Panahi, S. D{\"o}rscher, M. Baumert, E.-M. Richter, J. Kronj\"{a}ger, K. Bongs, and K. Sengstock,
Nature Phys. {\bf 4}, 496--501 (2008).

\bibitem{pe1}
C. Hamner, J. J. Chang, P. Engels, and M. A. Hoefer,
Phys.\ Rev.\ Lett. {\bf 106}, 065302 (2011).

\bibitem{pe2}
S. Middelkamp, J. J. Chang, C. Hamner, R. Carretero-Gonz{\'a}lez, P. G. Kevrekidis,
V. Achilleos, D. J. Frantzeskakis, P. Schmelcher, and P. Engels,
Phys.\ Lett.\ A {\bf 375}, 642--646 (2011).

\bibitem{pe3}
D. Yan, J. J. Chang, C. Hamner, P. G. Kevrekidis, P. Engels, V. Achilleos,
D. J. Frantzeskakis, R. Carretero-Gonz{\'a}lez, and P. Schmelcher,
Phys.\ Rev.\ A {\bf 84}, 053630 (2011).

\bibitem{azu}
A. {\'A}lvarez, J. Cuevas, F. R. Romero, C. Hamner, J. J. Chang, P. Engels,
P. G. Kevrekidis, and D. J. Frantzeskakis,
J. Phys. B: At. Mol. Opt. Phys. {\bf 46}, 065302 (2013).

\bibitem{pe4}
M. A. Hoefer, J. J. Chang, C. Hamner, and P. Engels,
Phys.\ Rev.\ A {\bf 84}, 041605(R) (2011).

\bibitem{pe5}
D. Yan, J. J. Chang, C. Hamner, M. Hoefer, P. G. Kevrekidis, P. Engels, 
V. Achilleos, D. J. Frantzeskakis, and J. Cuevas,
J.\ Phys.\ B: At.\ Mol.\ Opt.\ Phys. {\bf 45}, 115301 (2012).

\bibitem{spinor} T. M. Bersano, V. Gokhroo, M. A. Khamehchi, J. D'Ambroise, 
D. J. Frantzeskakis, P. Engels, and P. G. Kevrekidis, 
Phys. Rev. Lett. {\bf 120}, 063202 (2018).

\bibitem{pgkdjf} P. G. Kevrekidis and D. J. Frantzeskakis,
Reviews in Physics {\bf 1}, 140 (2016).

\bibitem{wenlong2} 
W. Wang, and P. G. Kevrekidis, Phys. Rev. E {\bf 95}, 032201 (2017).

\bibitem{wenlong3} 
E. G. Charalampidis, W. Wang, P. G. Kevrekidis, D. J. Frantzeskakis, and J. Cuevas-Maraver,
Phys. Rev. A {\bf 93}, 063623 (2016).

\bibitem{law} K. J. H. Law, P. G. Kevrekidis, and Laurette S. Tuckerman
Phys. Rev. Lett. {\bf 105}, 160405 (2010)

\bibitem{pola} M. Pola, J. Stockhofe, P. Schmelcher, and P. G. Kevrekidis
Phys. Rev. A {\bf 86}, 053601 (2012).


\bibitem{braz} V. A. Brazhnyi, and V. M. P{\'e}rez-Garc{\'i}a, 
Chaos, Solitons and Fractals {\bf 44}, 381--389 (2011). 

\bibitem{middel_pra} S. Middelkamp, P. G. Kevrekidis,
  D. J. Frantzeskakis, R. Carretero-Gonz{\'a}lez, and P. Schmelcher, 
Phys. Rev. A {\bf 82}, 013646 (2010).

\bibitem{jans} J. Stockhofe, P. G. Kevrekidis, D. J. Frantzeskakis, and P. Schmelcher,
J. Phys. B: At. Mol. Opt. Phys. {\bf 44}, 191003 (2011).


\end{thebibliography}
\end{document}